# Microscopic mechanisms of Strong Electron Scattering and Giant Anomalous Hall Effect in high-Curie-temperature Fe$_3$GaTe$_2$ van der Waals Films

Zhengxiao Li[1,2], Xin Lin[1,2], Yu Zou[3], Fanjie Tan[3], Wenliang Zhu,[3*] and Lijun Zhu[1,2**]

1. *State Key Laboratory of Semiconductor Physics and Chip Technologies, Institute of Semiconductors, Chinese Academy of Sciences, Beijing 100083, China*
2. *Center of Materials Science and Optoelectronics Engineering, University of Chinese Academy of Sciences, Beijing 100049, China*
3. *College of Physics and Information Technology, Shaanxi Normal University, Xi'an 710062, China*
\* wlzhu@snnu.edu.cn, \*\*ljzhu@semi.ac.cn

Van der Waals ferromagnet Fe$_3$GaTe$_2$ with room-temperature perpendicular magnetic anisotropy and strong anomalous Hall effect has attracted considerable interest for their potential in spintronics. However, the microscopic mechanisms and manipulation of the electron scattering and the anomalous Hall effect of Fe$_3$GaTe$_2$ have remained unsettled. Here, we demonstrate strong tuning of the electron scattering and anomalous Hall effect of pattern-defined Fe$_3$GaTe$_2$ Hall-bar devices with perpendicular magnetic anisotropy, high Curie temperature (340 K, as high as that of Fe$_3$GaTe$_2$ bulk), and giant anomalous Hall effect by varying the layer thickness and temperature. Temperature-dependent resistivity experiments reveal that the electron scattering of the high-quality Fe$_3$GaTe$_2$ is dominated by impurity scattering and phonon scattering, regardless of the thickness. Combined temperature- and thickness-dependent scaling analyses of the anomalous Hall resistivity reveal that the anomalous Hall effect of the Fe$_3$GaTe$_2$ is predominantly from the positive, temperature-independent skew-scattering contribution that competes with negative temperature-independent, side-jump contribution, and negative, temperature-dependent intrinsic Berry-curvature contribution. The intrinsic anomalous Hall conductivity decreases rapidly with increasing impurity scattering, which is consistent with the characteristic variation of intrinsic Hall conductivities in the dirty-metal regime. These findings advance the understanding of electron scattering and the anomalous Hall effect in van der Waals magnets and would benefit the application of the Fe$_3$GaTe$_2$ in spintronics.

Van der Waals magnets have received blooming interest in material science and spintronics for the advantages of easy dislocation, flexible stacking, strong tunability, and dangling-bond-free interfaces [1-17]. Among the various van der Waals magnets, Fe$_3$GaTe$_2$ is of particular interest for room-temperature ferromagnetism [18-24], perpendicular magnetic anisotropy, tunable chiral magnetic domains and skyrmions [21,23,24], efficient magnetization switching by spin-orbit torque [25-29], and flexible stacking of magnetic tunnel junctions with high tunneling magnetoresistance [30,31].

From the viewpoint of spintronics, a comprehensive understanding of the electron scattering and the anomalous Hall effect (AHE) is of key importance for applications of a spintronic material. For instance, electron scattering affects the generation [32-34] and relaxation [35,36] of spin current via spin-orbit coupling effects, and further spin-orbit torques exerted on a magnetic layer by a given spin current. The AHE is widely used as the indicator of the magnetic anisotropy and magnetization orientation in a variety of experiments (e.g., harmonic Hall voltages [32-37] and magnetization switching [25-29,38,39]). In general, electron scattering arises from scattering of conduction electrons by impurities, phonons, magnons, and disorder effects (Kondo effect [40,41], electron-electron interaction [42,43], etc.), which can be investigated from the temperature dependence of the longitudinal resistivity ($\rho_{xx}$). The AHE has the intrinsic Berry curvature contribution and extrinsic side-jump and skew scattering contributions. The strengths of the three AHE contributions can be potentially disentangled from the scaling of anomalous Hall resistivity ($\rho_{AH}$) with $\rho_{xx}$ for high-quality samples in which the scattering of conducting electrons is dominated by impurities and phonons [44-46]. In this case, $\rho_{AH}$ of a given sample measured from different temperatures follows [44,47]

$$\rho_{AH} = \alpha\rho_{xx0} + \beta\rho_{xx0}^2 + b\rho_{xx}^2, \quad (1)$$

where $b\rho_{xx}^2$ is the intrinsic contribution, and $\rho_{AH,extr} = \alpha\rho_{xx0} + \beta\rho_{xx0}^2$ is the sum of the extrinsic skew-scattering contribution ($\alpha\rho_{xx0}$) and side-jump contribution ($\beta\rho_{xx0}^2$) [44-47]. Here, $\alpha$ and $\beta$ are temperature-independent constants, $\rho_{xx0}$ is the residual resistivity due to static impurity scattering and thus independent of temperature, $b$ is the intrinsic anomalous Hall conductivity that is independent of temperature for many ferromagnets [44-47]. Equation (1) describes well the AHE scaling of highly ordered ferromagnets [44-47] but losses accuracy for highly disordered or poor-conductivity material systems in which the resistivity is dominated by electron-electron interactions [42,43], orbital two-channel Kondo effect [48], or hopping conduction [49].

So far, the microscopic mechanisms of the electron scattering and the AHE in Fe$_3$GaTe$_2$ films have remained unsettled despite the efforts on un-patterned Fe$_3$GaTe$_2$ bulk and few-layer flakes [19,22,50-53] in which the irregular shapes and spatially non-uniform current distribution would create large uncertainties in the quantitative estimation of $\rho_{AH}$ and $\rho_{xx}$ [22]. Accurate study of the microscopic mechanisms of the electron scattering and the AHE of Fe$_3$GaTe$_2$ requires the definition of large-area, uniform Fe$_3$GaTe$_2$ films with controlled thicknesses into complete Hall-bar devices via photolithography and etching, which is typically challenging. So far, the scaling law of Equation (1) and separation of the three different AHE contributions have never been achieved in any Fe$_3$GaTe$_2$ films. The thickness dependences of the electron scattering and the AHE have remained largely unexplored.

Here, we report strong tuning of the electron scattering and the AHE of patterned Fe$_3$GaTe$_2$ Hall-bar devices with perpendicular magnetic anisotropy by layer thickness ($t$) and temperature ($T$). From this ability, we demonstrate the microscopic mechanisms of the electron scattering and the AHE of high-quality Fe$_3$GaTe$_2$ films from the scaling of Eq. (1).

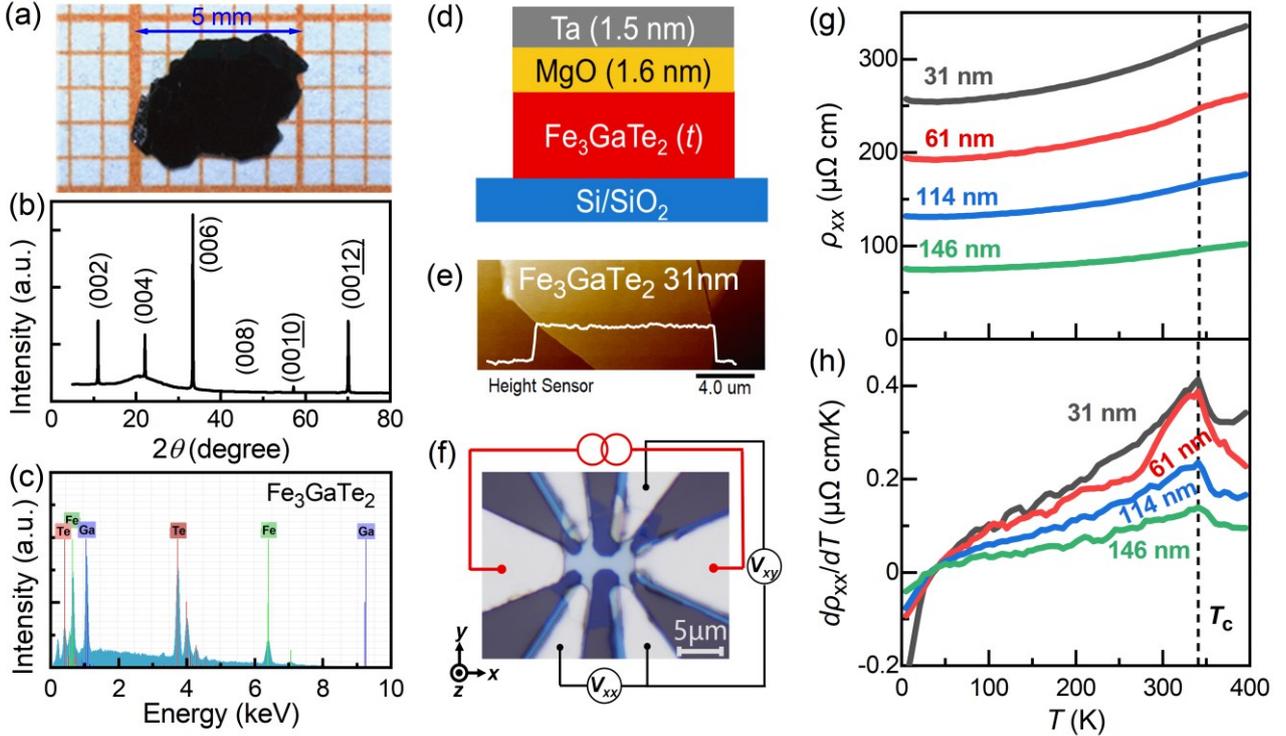

**Figure 1. Samples and electron scattering.** (a) Image of the as-grown $Fe_3GaTe_2$ single crystal that is placed on a millimeter grid. (b) X-ray diffraction $\theta$-$2\theta$ pattern of the $Fe_3GaTe_2$ single crystal, suggesting a (001) orientation. (c) Energy-dispersive x-ray spectroscopy pattern of the $Fe_3GaTe_2$ single crystal. (d) Schematic of the sample structure. (e) Atomic force microscopy image of the 31 nm $Fe_3GaTe_2$ layer. (f) Optical microscopy image of the two-cross Hall bar device and the measurement configuration. (g) Dependence on temperature ($T$) of the longitudinal resistivity ($\rho_{xx}$) for the $Fe_3GaTe_2$ samples with different thicknesses, suggesting a strong tuning of the resistivity via temperature and thickness. (h) Temperature dependence of $d\rho_{xx}/dT$, suggesting a Curie temperature ($T_c$) of ≈340 K for all the $Fe_3GaTe_2$ samples.

For this work, a high-quality $Fe_3GaTe_2$ single crystal was grown by the flux method (see Fig. 1a and the Experimental Section). The x-ray diffraction $\theta$-$2\theta$ pattern exhibits only (00$l$) diffraction peaks, confirming the (001) oriented single-crystalline texture (Fig. 1b). Energy-dispersive X-ray spectroscopy (EDS) analysis of the $Fe_3GaTe_2$ reveals an atomic ratio fairly close to the stoichiometric ratio of Fe:Ga:Te = 3:1:2 and the absence of any foreign elements within the experimental resolution (Fig. 1c). For transport study, four $Fe_3GaTe_2$ flakes with different layer thicknesses were mechanically exfoliated from the $Fe_3GaTe_2$ single crystal and then transferred onto oxidized silicon substrates (Si/SiO$_2$) in an argon glove box with the H$_2$O and O$_2$ pressures below 0.01 ppm. Subsequently, each sample was uniformly protected from oxidation by a sputter-deposited MgO (1.6 nm)/Ta (1.5 nm) bilayer (Fig. 1d), in which the Ta is expected to be fully oxidized after exposure in the atmosphere (see the electron energy loss spectroscopy evidence in Ref. [54]). Such MgO (1.6 nm)/Ta (1.5 nm) bilayer has been proven to protect very well many kinds of magnetic samples in the field of spintronics [32-39]. The transfer time of the samples from the glove box to the ultrahigh-vacuum sputtering system was managed to be less than 10 seconds to avoid potential oxidization. The layer thicknesses of the $Fe_3GaTe_2$ samples were then determined from atomic force microscopy measurements to be approximately 31 nm (Fig. 1e), 61 nm, 114 nm, and 146 nm, respectively. Each $Fe_3GaTe_2$/MgO/Ta stack was further patterned into two-cross Hall-bar devices with total length ($L$) of 10 μm and width ($W$) of 2 μm using photolithography and argon ion etching (Fig. 1f). Finally, electrical contacts of Ti 5 nm/Pt 150 nm were fabricated using photolithography, sputter-deposition, and liftoff. The transverse and longitudinal resistivities are calculated from the transverse and longitudinal voltages, $V_{xy}$ and $V_{xx}$, measured in a physical properties measurement system (PPMS-9T) while sourcing a small current ($I$) of 400 nA.

To provide insight into the electron scattering mechanism, we measure the resistivities of the $Fe_3GaTe_2$ with different layer thicknesses ($t_{FeGaTe}$) as a function of temperature (Fig. 1g). Upon cooling, $\rho_{xx}$ for each sample first decreases gradually towards a minimum resistivity (≈$\rho_{xx0}$) due to increasingly suppressed thermal phonon and magnon scattering and then increases only very slightly at very low temperatures (typically below 50 K). The low-temperature upturn is very weak in Fig. 1g but can be firmly identified from the small negative temperature derivate of resistivity ($d\rho_{xx}/dT$) in Fig. 1h. Similar low-temperature upturn has been previously observed in $Fe_3GaTe_2$ flakes [18,19,21,51,55] and bulks [28,50,52,53], $L1_0$-MnAl films [40], $L1_0$-MnGa films [41], $L2_1$-Co$_2$MnAl films [42], and amorphous FeTb films [43]. Such resistivity upturn has been typically attributed to the occurrence of disorder-induced Kondo effects (~ln$T$) or electron-electron interaction (~$T^{1/2}$). However, the resistivity upturns in the $Fe_3GaTe_2$ samples of this study are so weak that the exact mechanism cannot be distinguished reliably from the temperature dependence. The fact that the resistivity upturns of our $Fe_3GaTe_2$ devices are negligibly weak compared to those in previously reported Fe$_3$GaTe$_2$ flakes [18,19,21,51,55] and bulks [28,50,52,53] implies the enhanced quality of the Fe$_3$GaTe$_2$ samples in this work.

The minimum resistivity of the Fe$_3$GaTe$_2$ is tuned by 3.5 times from 254 μΩ cm to 74 μΩ cm as the layer thickness is increased from 31 nm to 146 nm, which indicates an enhancement of disorder scattering upon thinning of the Fe$_3$GaTe$_2$ layer. The strong dependence of the resistivity on the layer thickness at each fixed temperature is an intriguing observation due to some exotic mechanism yet to know. We note that strong thickness dependence of the resistivity is commonly seen in many magnetic layers, including amorphous rare-earth transition-metal alloys (e.g., FeTb [43]) to single-crystalline 3$d$ ferromagnets (e.g., Fe [47]). We infer that the role of the bulk and/or surface defects (such as atomic defects near the interfaces with the bottom SiO$_2$ or the top MgO layers) became more significant in the determination of electron scattering upon the decrease of the layer thickness. It is known that mechanical exfoliation and transfer of van der Waals materials typically introduce interlayer misalignment and intralayer stretching of atomic lattices [56], leading to significant strain and distortion that can function as bulk defects for electron transport. In contrast, interfacial scattering is unlikely to play any significant role in the determination of $\rho_{xx}$ of these thick, resistive Fe$_3$GaTe$_2$ because they should have a very short mean-free path. Moreover, the $d\rho_{xx}/dT$ data for each sample exhibit a broad peak likely due to the vanishing of a weak but non-zero magnon-electron scattering beyond the Curie temperature of approximately 340 K for all four samples with different thicknesses (Fig. 1h). However, the variation of the magnon scattering near the Curie temperature does not create a significant kirk in the temperature dependence of $\rho_{xx}$ in Fig. 1g, suggesting that the scattering of the electrons in the Fe$_3$GaTe$_2$ is dominated by impurities and phonons, but not magnons.

In Fig. 2a we plot the dependence on the perpendicular field ($H_z$) of the transverse resistivity ($\rho_{xy}$) for the Fe$_3$GaTe$_2$ samples with different thicknesses at temperatures ranging from 5 to 360 K. The switching of the Fe$_3$GaTe$_2$, regardless of the thickness and temperature, exhibits a gradual multistate process rather than a sharp up-down switching, which suggests a non-uniform multidomain switching process. More specifically, the gradual switching of the Fe$_3$GaTe$_2$ devices suggests a large distribution of anti-domain nucleation fields and/or domain wall depinning fields that are sensitive to defects. Such defects could be formed randomly during the crystal growth and device fabrication (e.g., mechanical exfoliation, transfer, sputtering deposition of the protective bilayer, photolithography, or etching).

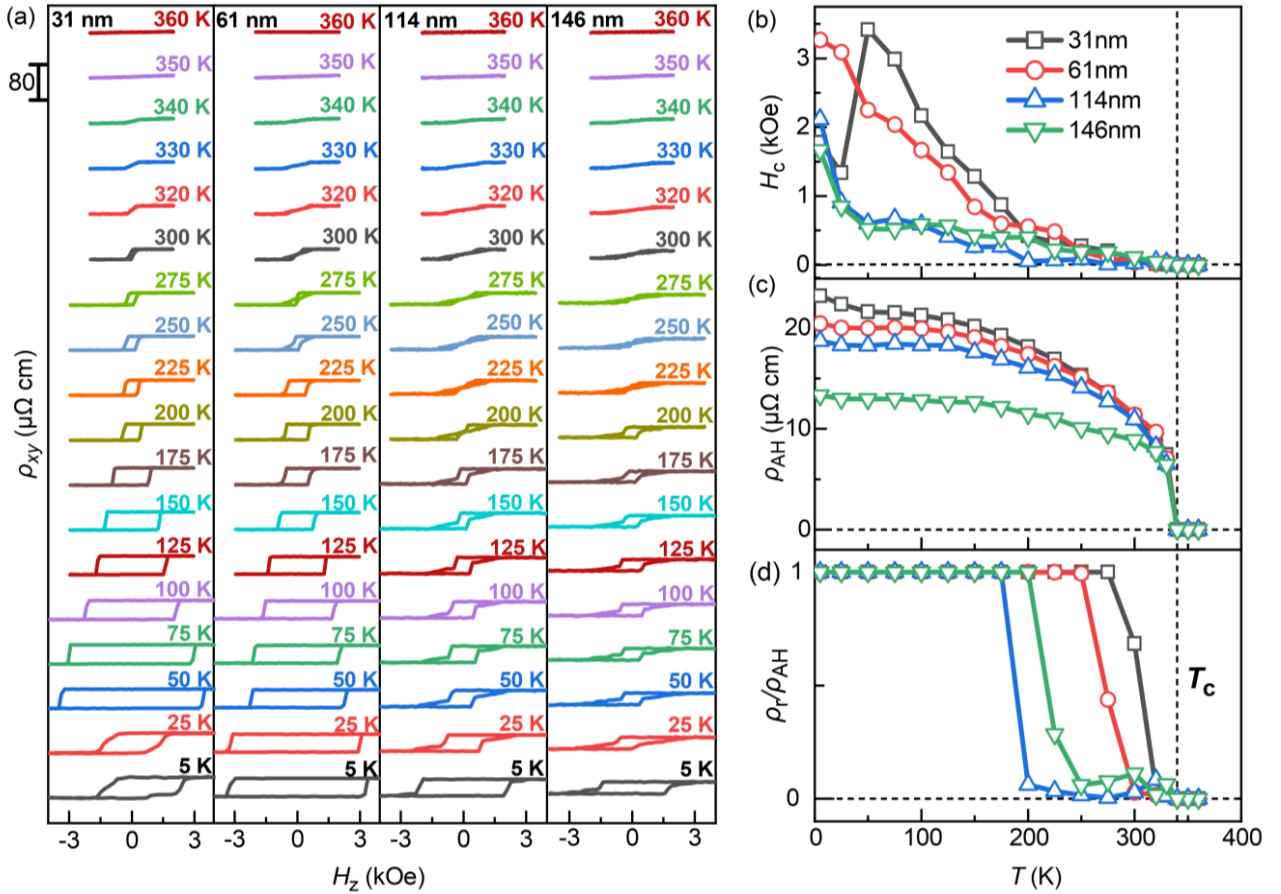

**Figure 2. Temperature-dependent measurement o-f the anomalous Hall effect.** (a) Dependence on the perpendicular field ($H_z$) of the transverse resistivity ($\rho_{xy}$) for the Fe$_3$GaTe$_2$ samples with different thicknesses at temperatures ranging from 5 to 360 K. Temperature dependence of (b) the coercivity ($H_c$), (c) anomalous Hall resistivity ($\rho_{AH}$) and (d) the remanent ratio ($\rho_r/\rho_{AH}$) for the Fe$_3$GaTe$_2$ Hall bar devices. The vanishing temperature for $H_c$, $\rho_{AH}$, and $\rho_r/\rho_{AH}$ indicates the Curie temperature ($T_c$) of 340 K for the Fe$_3$GaTe$_2$ samples with different thicknesses.

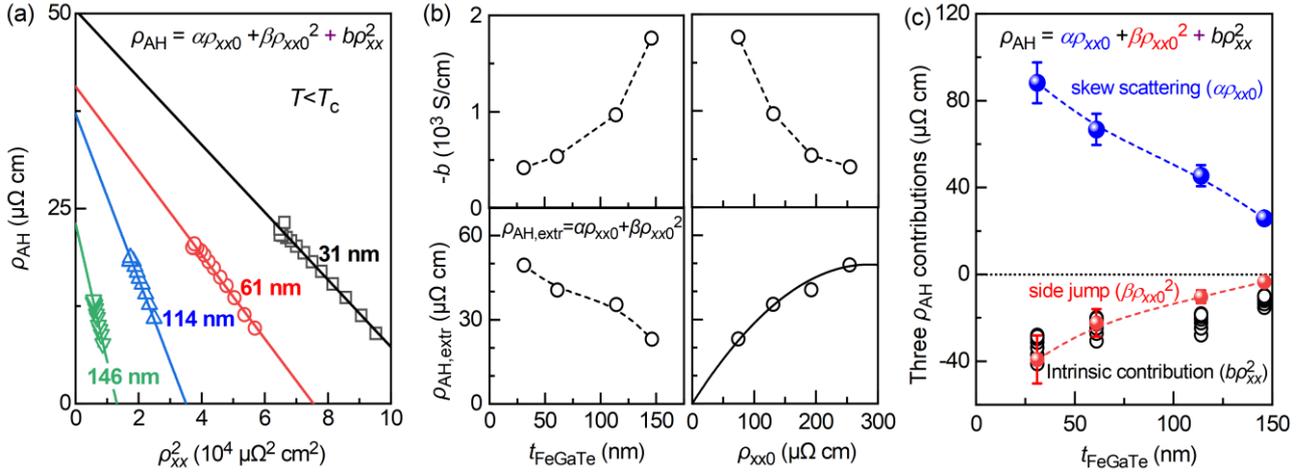

**Figure 3. Scaling law of the anomalous Hall effect.** (a) Dependence of $\rho_{AH}$ on $\rho_{xx}^2$. The solid lines represent the linear fits of the data of each sample as measured at different temperatures following Eq. (1). (b) Dependences of the sum extrinsic anomalous Hall resistivity ($\rho_{AH,extr}$) and the intrinsic anomalous Hall conductivity ($b$) on the layer thickness ($t_{FeGaTe}$) and residual resistivity ($\rho_{xx0}$), the solid curve represents the quadratic fit. (c) Skew scattering (blue dots), side jump (red dots) and intrinsic contributions (black circles) to the anomalous Hall resistivity of the Fe$_3$GaTe$_2$ with different thicknesses.

It is interesting to note that, while most of the switching hysteresis loops indicate essentially the same up-to-down and down-to-up switching fields, the 31 nm Fe$_3$GaTe$_2$ sample exhibits an exchange bias fields of 29 Oe at 25 K and 566 Oe at 5 K. The exchange bias in Fe$_3$GaTe$_2$ at low temperatures is an interesting observation that requires future efforts to fully understand. Note that low-temperature exchange bias has also been observed previously in oxidized Fe$_3$GaTe$_2$ [57,58] as well as in non-oxidized magnetic materials [59,60]. Since all the Fe$_3$GaTe$_2$ samples in this work are carefully protected in the same way, the absence of the low-temperature exchange bias in our Fe$_3$GaTe$_2$ samples with large thicknesses should have excluded oxidization as the cause of the low-temperature exchange bias. This conclusion is consistent with the minimal resistivity upturn in the resistivity in Fig. 1g since significant oxidation should greatly increase the low-resistivity upturn. Instead, the occurrence of the exchange bias only at the smallest thickness could be related to magnetic disorder induced by strain and distortion from mechanical exfoliation and transfer of the van der Waals material.

From the $\rho_{xy}$-$H_z$ data, the perpendicular coercivity ($H_c$), the anomalous Hall resistivity ($\rho_{AH}$), the remanent ratio ($\rho_r/\rho_{AH}$) are determined and summarized in Fig. 2b-d. Here, $H_c$ is the perpendicular switching magnetic field. $\rho_{AH}$ is determined from the $\rho_{xy}$-$H_z$ data in the high-field regime after the subtraction of the linear-in-$H_z$ ordinary Hall resistivity contribution. $\rho_r$ is the remanent Hall resistivity at zero magnetic field. $H_c$, $\rho_{AH}$, and $\rho_r/\rho_{AH}$ for the Fe$_3$GaTe$_2$ remain zero at temperatures beyond ≥340 K but emerge at lower temperatures, which is consistent with the indication of the Curie temperature of 340 K by the electron scattering (Fig. 1g,h). The Curie temperature of 340 K is the highest among the reported Fe$_3$GaTe$_2$ thin-film samples. For the three Fe$_3$GaTe$_2$ samples with large thicknesses of 61 nm, 114 nm, and 146 nm, $H_c$ increases monotonically upon cooling at low temperatures (Fig. 2b), which can be attributed to the suppression of thermally-assisted anti-domain nucleation [61] and/or domain wall depinning at low temperatures [62]. In contrast, $H_c$ for the 31 nm Fe$_3$GaTe$_2$ also shows an abrupt drop at 25 K and then an increase at 5 K, which are reaffirmed by repeated measurements. The non-monotonic dependences of $H_c$ and the presence of the significant exchange bias at 5-25 K suggest an important change in the microscopic switching process of the 31 nm Fe$_3$GaTe$_2$.

As shown in Fig. 2c, $\rho_{AH}$ for the Fe$_3$GaTe$_2$ increases monotonically upon cooling in the entire temperature range below the Curie temperature. This temperature dependence is distinct from that of most 3$d$ metallic magnets (Fe, Co, Ni, MnGa, etc.)[44-46] in which $\rho_{AH}$ typically drops upon cooling. In analogy to the longitudinal resistivity (Fig.1g), $\rho_{AH}$ at each fixed temperature increases as the thickness decreases, which is contrary to the previous report of the decrease of $\rho_{AH}$ with layer thickness for few-layer Fe$_3$GaTe$_2$ flakes [21]. The temperature dependence of the remanent ratio $\rho_r/\rho_{AH}$ also varies strongly with the thickness (Fig. 2d). $\rho_r/\rho_{AH}$ for the 31 nm Fe$_3$GaTe$_2$ increases very rapidly to unity at 275 K, while increasingly lower temperature as the thickness increases (i.e., 250 K for 61 nm, 175 K for 114 nm, and 200 K for 146 nm). This effect can be understood by the improved magnetic uniformity of the Fe$_3$GaTe$_2$ at small thicknesses. We note that the thickness dependences of gradual switching, coercivity, and $\rho_r/\rho_{AH}$ are not intrinsic dimensional confinement effects but arise simply from the non-uniform nucleation and/or depinning fields in the presence of defects formed randomly during the crystal growth and device fabrication.

To unveil the microscopic mechanisms of the anomalous Hall effect of the Fe$_3$GaTe$_2$, we now discuss the scaling of $\rho_{AH}$ with $\rho_{xx}$. The scaling analysis can disentangle the intrinsic [63] and extrinsic contributions [64,65] of the anomalous Hall resistivity of magnetic materials. As plotted in Fig. 3a, the anomalous Hall resistivity of the Fe$_3$GaTe$_2$ scales linearly with $\rho_{xx}^2$ in the wide temperature below the Curie temperature (5-320 K), which is well consistent with Eq. (1). This result suggests that the anomalous Hall resistivity of the Fe$_3$GaTe$_2$ is predominantly governed by intrinsic Berry curvature contribution and the temperature-independent impurity-induced extrinsic contributions, while the temperature-dependent electron scattering by phonons, magnons, and disorder effects only have a minor contribution to the extrinsic anomalous Hall resistivity. Note that this result represents the first observation of the good agreement of the AHE data with Eq. (1), suggesting the high quality of the $Fe_3GaTe_2$ device studied in this work.

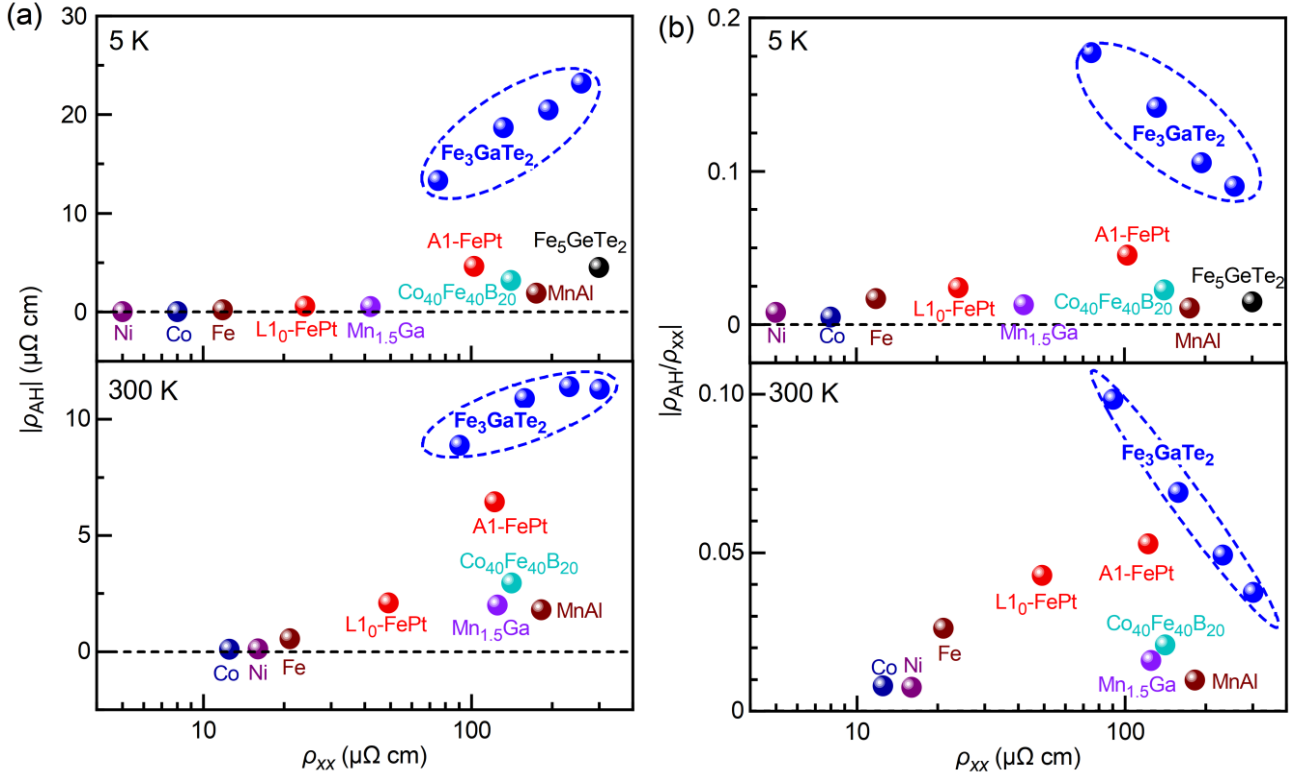

**Figure 4. Giant anomalous Hall effect.** Comparison of (a) the anomalous Hall resistivity $|\rho_{AH}|$, (b) the anomalous Hall angle $|\rho_{AH}/\rho_{xx}|$, and the longitudinal resistivity ($\rho_{xx}$) at 5 K and 300 K for the $Fe_3GaTe_2$ and other representative magnetic films (Fe [47], Co [45], Ni [46], $L1_0$-FePt [67], A1-FePt [67] $Co_{40}Fe_{40}B_{20}$ [72], $Mn_{1.5}Ga$ [44], MnAl [48], and $Fe_5GeTe_2$ [73]), highlighting the giant AHE of the $Fe_3GaTe_2$.

The slope and intercept of the linear fits of the data in Fig. 3a yield the values of the intrinsic anomalous Hall conductivity $b$ and the sum extrinsic anomalous Hall resistivity $\rho_{AH,extr} = \alpha\rho_{xx0} + \beta\rho_{xx0}^2$, respectively. As shown in Fig. 3b, as the layer thickness increases, $b$ remains negative and increases monotonically in magnitude by a factor of 4 from -415 S/cm at $t_{FeGaTe}$ = 31 nm to -1771 S/cm at $t_{FeGaTe}$= 146 nm. In contrast, $\rho_{AH,extr}$ is positive and reduces by more than a factor of 2 upon the increase of the layer thickness, from 50 μΩ cm at $t_{FeGaTe}$ =31 nm to 23 μΩ cm at $t_{FeGaTe}$ =146 nm. The experimental value of the intrinsic anomalous Hall conductivity agrees reasonably in sign and magnitude with the previous first-principles calculations [66]. Since a striking characteristic of the intrinsic Hall conductivities is the rapid reduction with increasing electron scattering in the dirty-metal regime (see the intrinsic AHE in Mn-Ga [44], Co [45], FePt [67], and CoPt [68], and the intrinsic spin Hall effect in Pt-X alloys [32,36,69-71]), we attribute the thickness dependence of the intrinsic anomalous Hall conductivity of the $Fe_3GaTe_2$ to the variation of the resistivity with the layer thickness (Fig. 1g). This mechanism is further verified by the decrease of $b$ with $\rho_{xx0}$ in Fig. 3b. From the fits of the $\rho_{AH,extr}$ vs $\rho_{xx0}$ data, we determine that $\alpha$ = 0.35±0.04 and $\beta$ = (-6.07±1.71)×$10^{-4}$/μΩ cm and further separate the skew scattering $\alpha\rho_{xx0}$ and side jump contributions $\beta\rho_{xx0}^2$.

In Fig. 3c, we plot the intrinsic and extrinsic anomalous Hall resistivities of the $Fe_3GaTe_2$ as a function of layer thickness. It is seen that the anomalous Hall resistivities of the $Fe_3GaTe_2$ are dominated by the positive skew scattering contribution, while the negative intrinsic and side-jump contributions are significantly smaller. As temperature increases, the dominant skew-scattering anomalous Hall resistivity is increasingly canceled by the intrinsic contribution that increases as the temperature increases, which explains the positive sign and magnitude reduction of $\rho_{AH}$ upon warming up in Fig. 2c.

Finally, we note that the anomalous Hall resistivity of the $Fe_3GaTe_2$ is giant compared to that of FePt [67] and the 3$d$ magnets Fe [47], Co [45], Ni [46], $Co_{40}Fe_{40}B_{20}$ [72], $Mn_{1.5}Ga$ [44], MnAl [48], and $Fe_5GeTe_2$ [73] at low temperature and room temperature (Fig. 4a). As shown in Fig. 4b, the $Fe_3GaTe_2$ also has very large anomalous Hall angle ($|\rho_{AH}/\rho_{xx}|$), especially the 146 nm $Fe_3GaTe_2$ exhibits high $\rho_{AH}/\rho_{xx}$ of 0.18 at 5 K and 0.1 at 300 K, which are highly preferred for sensor applications in a wide temperature range.

In conclusion, we have fabricated high-quality $Fe_3GaTe_2$ Hall-bar devices with perpendicular magnetic anisotropy, high Curie temperature (340 K, which is as high as that of $Fe_3GaTe_2$ bulk [18]), and giant anomalous Hall effect. We demonstrate that the electron scattering and the anomalous Hall resistivity of the high-quality $Fe_3GaTe_2$ can be tuned strongly by the layer thickness and temperature. We find that the electron scattering is dominated by impurity scattering and phonon scattering. The anomalous Hall resistivity of the $Fe_3GaTe_2$ exhibits a well-defined scaling of $\rho_{AH} = \alpha\rho_{xx0} + \beta\rho_{xx0}^2 + b\rho_{xx}^2$ in the wide temperature below the Curie temperature. From the scaling analysis, we determine that the anomalous Hall effect of the $Fe_3GaTe_2$ is dominated by a positive, temperature-independent skew scattering that competes with the negative side-jump and intrinsic contributions. The intrinsic anomalous Hall conductivity decreases with increasing impurity scattering, which is consistent with the characteristic variation of intrinsic Hall conductivities in the dirty-metal regime. These findings advance the understanding of electron scattering and the anomalous Hall effects in van der Waals magnets and would benefit the application of the $Fe_3GaTe_2$ in spintronics.

**Experimental Section**
**Single crystal growth**: High-quality single crystals of $Fe_3GaTe_2$ were grown by the flux method. The mixture of Fe powder (99.98%), Ga ingots (99.99%), and Te ingots (99.999%) was placed into an alumina crucible with a molar ratio of 1:1:2 and sealed in an evacuated quartz tube. The quartz tube was heated to 1000°C, and held there for 48 h. The temperature was then cooled to 880°C in 1 h, and further cooled to 780°C at a rate of 1°C/h. Finally, the excess Te flux was removed with a centrifuge. The obtained single crystals are of thin-plate shape.

**Sample characterizations**: The crystal structure of as-prepared $Fe_3GaTe_2$ was characterized using a Bruker diffractometer with Cu K$\alpha$ radiation at room temperature. The energy-dispersive x-ray spectroscopy analysis of $Fe_3GaTe_2$ was performed in a field emission electron microscope (Nova NanoSEM450). Transport measurements are performed at different temperatures in a Quantum Design physical properties measurement system (PPMS-9T). The layer thickness of the $Fe_3GaTe_2$ is measured after coating of the protective layer using atomic force microscopy (Bruker Dimension Icon).

**Acknowledgment**
We thank Changmin Xiong, Xintian Wu, and Tianshuo Shi for their help with the use of the PPMS-9T system. This work is supported partly by the National Key Research and Development Program of China (2022YFA1204000), the Beijing National Natural Science Foundation (Z230006), and by the National Natural Science Foundation of China (12304155, 12274405, and 12204297).